\documentclass[aps,prl,final,twocolumn,showpacs,showkeys,floatfix]{revtex4}

\usepackage{amsmath}
\usepackage{graphicx}
\usepackage{amssymb,amsfonts}
\usepackage{bm}
\usepackage{dcolumn}
\usepackage{natbib}
\usepackage[usenames]{color}
\usepackage[dvipdfm,colorlinks=true,pdfborder={0 0 0}]{hyperref}
\hypersetup{%
  pdftitle={Intermittent magnetic field excitation by a turbulent
flow of liquid sodium},
  pdfauthor={Mark D. Nornberg},
  pdfstartview={FitH},
  linkcolor=MidnightBlue,
  citecolor=BlueViolet,
  urlcolor=Plum,
  anchorcolor=Plum
}

\newcommand{\pder}[3][]{\frac{\partial^{#1} #2}{\partial #3^{#1}}}
\newcommand{\del}{\boldsymbol{\nabla}}

\newcommand{\cross}{\times}

\newcommand{\curl}[1]{\del\cross{ \bf #1 }}

\begin{document}

\title{Intermittent magnetic field excitation by a turbulent
flow of liquid sodium}

\author{M.~D.~Nornberg}
\author{E.~J.~Spence}
\author{R.~D.~Kendrick}
\author{C.~M.~Jacobson}
\author{C.~B.~Forest}
\email{cbforest@wisc.edu}

\affiliation{Department of Physics\\ 
  University of Wisconsin-Madison\\
  1150 University Ave.\\ 
  Madison, WI 53706}

\date{\today}

\begin{abstract}
The magnetic field measured in the Madison Dynamo Experiment shows
intermittent periods of growth when an axial magnetic field is
applied. The geometry of the intermittent field is consistent with the
fastest growing magnetic eigenmode predicted by kinematic dynamo
theory using a laminar model of the mean flow. Though the eigenmodes
of the mean flow are decaying, it is postulated that turbulent
fluctuations of the velocity field change the flow geometry such that
the eigenmode growth rate is temporarily positive.  Therefore, it is
expected that a characteristic of the onset of a turbulent dynamo is
magnetic intermittency.
\end{abstract}

\keywords{magnetohydrodynamics, MHD, dynamo, turbulence,
  intermittency, Madison Dynamo Experiment, liquid sodium}
\pacs{47.65.+a, 91.25.Cw}

\maketitle

Determining the onset conditions for magnetic field growth in
magnetohydrodynamics is fundamental to understanding how astrophysical
dynamos such as the Earth, the Sun, and the galaxy self-generate
magnetic fields. These onset conditions are now being studied in
laboratory experiments using flows of liquid
sodium~\cite{Gailitis.RMP.2002}. The conditions required for
generating a dynamo can be determined by solving the magnetic
induction equation
\begin{equation}
\pder{\mathbf{B}}{t} = Rm \curl{V \times B} + \nabla^2 \mathbf{B},
\label{eq:induction}
\end{equation}
where $\mathbf{B}$ is the magnetic field, $\mathbf{V}$ is the velocity
field scaled by a characteristic speed $V_0$, and the time is scaled
to the resistive diffusion time $\tau_\sigma = \mu_0 \sigma L^2$. The
magnetic Reynolds number is $Rm=\mu_0\sigma L V_0$, where $\sigma$ is
the conductivity of the fluid and $L$ is a characteristic scale
length~\cite{Moffatt}. In the kinematic approximation, the velocity
field is assumed to be a prescribed flow (either the flow is
stationary or its time dependence is specified) and Lorentz forces due
to the magnetic field are neglected. Equation~\ref{eq:induction} is
then linear in $\mathbf{B}$ and can be solved as an eigenvalue
problem. Several different types of stationary, helical flows have
been shown theoretically to be kinematic
dynamos~\cite{Ponomarenko.JAMTP.1970, Gubbins.PRTSLA.1973,
Busse.GJRAS.1975, Dudley.PRSLA.1989}, which in turn has lead to the
design of current dynamo experiments. For particular flows, the
kinematic model predicts a critical value of the magnetic Reynolds
number, $Rm_{\rm crit}$, above which the magnetic field becomes
linearly unstable, {\em i.e.\ }for $Rm>Rm_{\rm crit}$ a small seed
magnetic field will grow exponentially in time. The dynamo onset
conditions have been tested in helical pipe-flow experiments at
Riga~\cite{Gailitis.PRL.2000,Gailitis.PRL.2001} and
Karlsruhe~\cite{Stieglitz.PF.2001,Muller.JFM.2004}. Both experiments
generated magnetic fields at a value of $Rm_{\rm crit}$ consistent
with predictions from the kinematic theory.

\begin{figure}
\centerline{\includegraphics{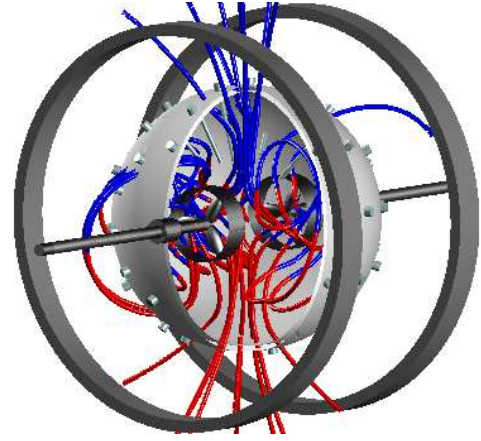}}
\caption{A schematic of the Madison Dynamo Experiment
  with superimposed magnetic field lines of the theoretically
  predicted dominant magnetic eigenmode.}
\label{fig:schematic}
\end{figure}

Fluids and plasmas such as the Earth's liquid core, the solar
convection zone, and liquid metals are turbulent under the conditions
required for a dynamo. The Riga and Karlsruhe experiments use highly
constrained flows to create the helical geometry necessary for
magnetic field generation. Astrophysical dynamos, however, are often
generated by flows in simply-connected geometries which do not exhibit
the scale separation between the large-scale magnetic field and the
fluctuating part of the velocity field employed in the Riga and
Karlsruhe experiments. This discrepancy has prompted the creation of
several experiments to investigate the dynamo onset conditions in
simply-connected flows with fully developed
turbulence~\cite{Forest.MHD.2002, Petrelis.PRL.2003,
Lathrop.PPCF.2001}.

The threshold for the dynamo instability is not expected to be the
smooth transition from decaying to growing magnetic fields described
by laminar kinematic theory~\cite{O'Connell.Cargese.2000}. Large-scale
eddies can cause the instantaneous flow to differ significantly from
the mean flow. The growth rate of the magnetic field is highly
sensitive to the flow geometry, and so the instantaneous flow may
occasionally satisfy $Rm>Rm_{\rm crit}$ while the mean flow does
not. The fastest-growing global magnetic eigenmode would then
fluctuate between growing and decaying states. The threshold of
magnetic field growth therefore has a range characterized by
intermittent bursts of magnetic field growth. In this Letter the
observation of an intermittently excited magnetic field in a
simply-connected, turbulent flow of liquid sodium is reported. The
structure of the excited field is consistent with the largest growing
magnetic eigenmode predicted from a laminar kinematic model of the
mean flow.

\begin{figure}
\includegraphics{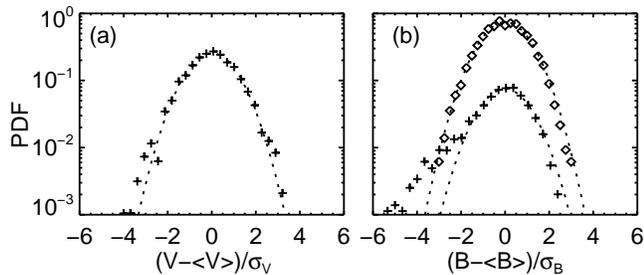}
\caption{(a) PDF constructed from LDV measurements of $v_\phi$ in the
  water model of the experiment ($+$) with Gaussian fit (dotted
  line). The difference between $v_\phi$ and its mean is scaled to the
  standard deviation $\sigma_V$. (b) PDF constructed from $B_r$
  measurements in the sodium experiment near the equator ($+$)
  and near the symmetry axis ($\diamond$) with Gaussian fit (dotted
  line). The difference between $B_r$ and its mean is scaled to the
  standard deviation $\sigma_B$.}
\label{fig:PDFs}
\end{figure}

The Madison Dynamo Experiment [Fig.~\ref{fig:schematic}] is
a 1~m diameter stainless steel sphere filled with liquid
sodium~\cite{Nornberg.PP.2006}.  Results presented in this Letter are
from a turbulent two-vortex flow, similar to the flows described in
\cite{Dudley.PRSLA.1989}, created by two counter-rotating
impellers. The impellers are each driven by 75~kW motors and can
achieve rotation rates up to 25~Hz, corresponding to $Rm_{\rm tip} =
\mu_0 \sigma L V_{tip} = 150$ based on the impeller tip speed.

The mean flow is designed to generate growing magnetic fields
according to a laminar kinematic dynamo model~\cite{Forest.MHD.2002}.
According to the kinematic eigenvalue calculations, for sufficiently
large $Rm$ the experimental flow is expected to excite a dipole
magnetic field oriented transverse to the symmetry axis
[Fig.~\ref{fig:schematic}].  An array of 74 temperature-compensated
Hall probes on the surface of the sphere provides measurements of the
instantaneous multipole structure of the magnetic field induced by
currents in the liquid sodium. The Hall probes are capable of
resolving changes in the magnetic field down to 0.3~G with a maximum
range of $\pm 170$~G. A pair of magnetic field coils coaxial with the
axis of rotation are used to apply a nearly-uniform 50~G axial field.
This seed field brings the field induced by the flow above the noise
level of the Hall probes. The applied field is sufficiently small that
the strength of the Lorentz force is about 1\% of fluid inertial
forces.

\begin{figure}
\centerline{\includegraphics{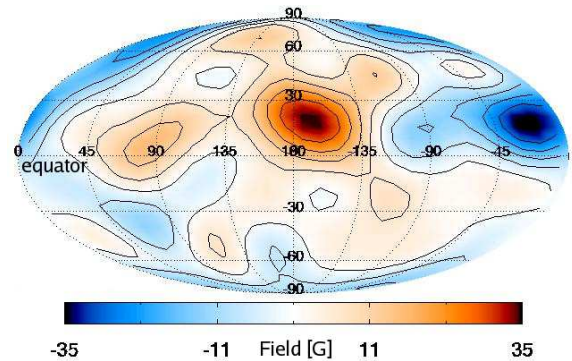}}
\caption{Contours of $B_r(\theta,\phi)$ measured on the
surface of the sphere. The applied field is subtracted from the
measurements. This snapshot of the measured field corresponds to an
induced dipole field transverse to the drive shaft axis.}
\label{fig:event_structure}
\end{figure}

The velocity field of an identical-scale water model of the experiment
is measured using Laser Doppler Velocimetry
(LDV)~\cite{Nornberg.PP.2006}. The velocity measurements have Gaussian
probability distribution functions (PDF) [Fig.~\ref{fig:PDFs}(a)] as
expected from a stationary turbulent flow according to the central
limit theorem~\cite{Tennekes_and_Lumley}.  The magnetic fluctuations
measured near the axis of symmetry of the sodium experiment also have
Gaussian distributions [Fig.~\ref{fig:PDFs}(b)]. In addition to these
normally-distributed fluctuations, there are intermittent,
large-amplitude magnetic bursts observed on probes near the equator of
the experiment. The magnetic field during a burst has the spatial
structure expected from the least-damped magnetic eigenmode from
kinematic theory [Fig.~\ref{fig:event_structure}]. The orientation of
the transverse dipole is random for each burst so that the time-averaged
induced field is axisymmetric.

\begin{figure}[h]
\centerline{\includegraphics{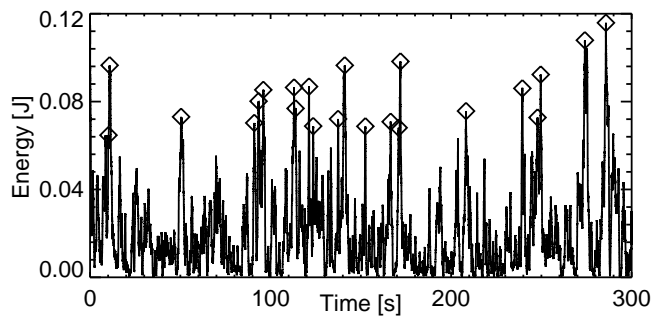}}
\caption{Time series of the energy in the transverse dipole field for
  an impeller rotation rate of 10~Hz. The diamonds mark the peak of a
  burst where the energy exceeds $50\%$ of its maximum value.}
\label{fig:timeseries}
\end{figure}

\begin{table*}
\caption{The magnetic Reynolds number $Rm$ based on the maximum speed
  in measured flows, duration of the measurement $T$, number of bursts
  $N_b$, average burst rate $f_b$, burst width $\tau_b$, growth rate
  $\lambda_b$, estimate of the overall fraction of time the flow is
  bursting $n_d$, mean energy $\left<E\right>$, and standard
  deviation of the energy $\sigma_E$ as a function of the
  rotation rate of the impellers $f_{\rm tip}$.}
\begin{ruledtabular}
\begin{tabular}{dcccdddddd}
\multicolumn{1}{c}{$f_{\rm tip}$ [Hz]} &
\multicolumn{1}{c}{$Rm$} &
\multicolumn{1}{c}{$T$ [s]} &
\multicolumn{1}{c}{$N_b$} &
\multicolumn{1}{c}{$f_b$ [s$^{-1}$]} &
\multicolumn{1}{c}{$n_d$ [\%]} &
\multicolumn{1}{c}{$\tau_b$ [s]} &
\multicolumn{1}{c}{$\lambda_b$ [s$^{-1}$]} &
\multicolumn{1}{c}{$\left<E\right>$ [mJ]} &
\multicolumn{1}{c}{$\sigma_E$ [mJ]}\\
\hline
3.3  & 14 & 300 &  5 & 0.017 & 6.7 & 3.99 & 0.17 &   2 &  2 \\
6.7  & 22 & 300 &  9 & 0.030 & 7.5 & 2.50 & 0.30 &   9 &  8 \\
10.0 & 28 & 300 & 22 & 0.070 & 6.1 & 0.83 & 1.12 &  21 & 20 \\
13.3 & 35 & 300 & 38 & 0.127 & 7.3 & 0.58 & 1.62 &  48 & 43 \\
16.7 & 42 & 300 & 37 & 0.123 & 6.3 & 0.51 & 2.22 &  78 & 76 \\
20.0 & 49 & 100 & 15 & 0.150 & 5.4 & 0.36 & 2.93 & 111 & 98 \\
\end{tabular}
\end{ruledtabular}
\label{tab:burst}
\end{table*}

\begin{figure}
\centerline{\includegraphics{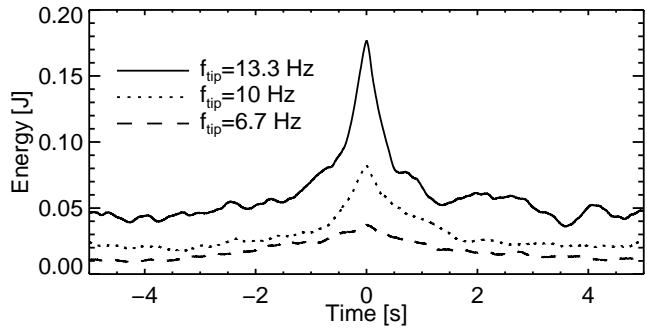}}
\caption{The ensemble average of bursts from three time series. The
  averaged burst is used to calculate the growth rate and burst width
  in Tab.~\ref{tab:burst}.}
\label{fig:condavg}
\end{figure}

The bursts are ensemble averaged to determine typical
characteristics. A burst is defined to occur when the energy in the
transverse dipole field exceeds a certain threshold. For this
analysis, the threshold is 50\% of the maximum energy of the time
series [Fig.~\ref{fig:timeseries}]. This threshold is sufficiently
small to capture a large number of bursts yet significantly larger
than the mean energy (about two standard deviations above the mean
energy for each time series). The bursts are averaged together and the
growth rate is determined by an exponential curve fit
[Fig.~\ref{fig:condavg}]. The results for various impeller rotation
rates are reported in Tab.~\ref{tab:burst}. It should be noted that
the strength of the fluctuations in the field is at most equal to the
on-axis applied field strength of 50\,G, hence the Lorentz force due
to the fluctuations is weak compared to inertial forces.

There are several possible mechanisms for the excitation of the
transverse dipole field. Velocity field fluctuations are large, with
$\widetilde{V}/\left<V\right> \approx 0.5$ as determined from LDV
measurements. These large fluctuations cause the peak flow speed to
vary, which can be interpreted as variation in $Rm$. Fluctuations at
the largest scales can also cause changes in the shape of the flow
leading to variation of $Rm_{\rm crit}$. The statistics of the
small-scale fluctuations could change, also contributing to variation
of $Rm_{\rm crit}$. If the kinetic helicity of the small-scale eddies
becomes sufficiently strong, the net current generation could give
rise to the observed magnetic field
bursts~\cite{Spence.PRL.2006,Cattaneo.PF.2005}. Regardless of scale,
subtle changes in the flow can adjust the instability threshold.

\begin{figure}
\centerline{\includegraphics{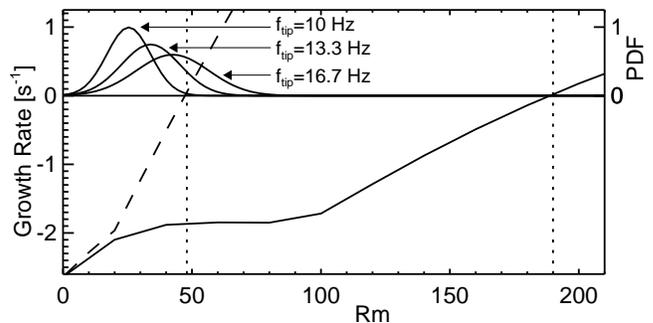}}
\caption{Kinematic growth rate versus $Rm$ for the mean flow measured
  in the water experiment (solid) and an optimized flow (dashed). The
  vertical lines identify $Rm_{\rm crit}$ for each case. The PDFs of
  $Rm$ for flows with three different impeller rotation rates are
  shown to demonstrate the increasing overlap of the ranges of $Rm$
  and $Rm_{\rm crit}$.}
\label{fig:growth_rate}
\end{figure}

To estimate the likelihood the flow is self-exciting, the kinematic
growth rate from the velocity field measured in the water experiment
is calculated [Fig.~\ref{fig:growth_rate}]. The solid line shows
the results of the calculation using the mean flow, whereas the dashed
line shows the growth rate calculated for an optimized flow geometry
similar to and within the fluctuation levels of the measured
flow~\cite{O'Connell.Cargese.2000}. The region between the growth
rates of the mean and optimized flows indicates the range of possible
eigenmode growth rates for an instantaneous realization of the flow
and the resulting variation in $Rm_{\rm crit}$.

The velocity fluctuations also contribute to variations in the
instantaneous maximum speed of the flow, thereby creating a range of
$Rm$. Probability distributions of $Rm$ constructed from the measured
velocity fluctuations for three different impeller rotation rates are
plotted in Fig.~\ref{fig:growth_rate}. It is expected that a greater
overlap of the PDF of $Rm$ with the range of $Rm_{\rm crit}$ will
result in magnetic field excitations with greater frequency and
strength. The duration of each excitation is expected to decrease
since the correlation time of the velocity fluctuations scales as
$\tau_c \sim \ell / V_\ell$ where $\ell$ is the eddy scale length and
$V_\ell$ is the characteristic speed of the eddy. For example, LDV
measurements from the water model give $\tau_c= 80 \pm 20$ ms for the
$f_{\rm tip}=16.7$\,Hz flow, consistent with eddies of size
$\ell=0.25$\,m and speed $V_\ell=3\,{\rm m}/{\rm s}$. The proportion
of time that the magnetic field is bursting is estimated to be $n_d =
f_b \tau_b$, where $f_b$ is the average frequency of the bursts and
$\tau_b$ is the width of the conditionally-averaged burst at
half-maximum. The data in Tab.~\ref{tab:burst} show that the
proportion of time the flow is bursting stays relatively constant
between 5--8\%.

\begin{figure}
\centerline{\includegraphics{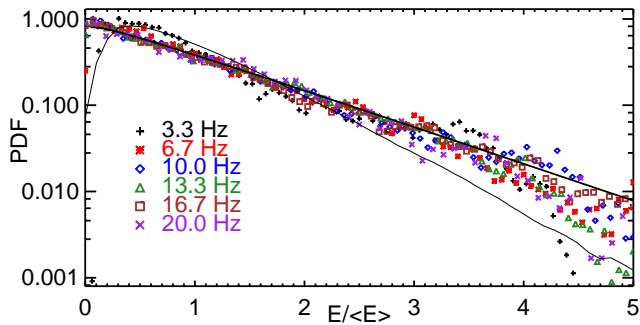}}
\caption{The PDF of the energy in the transverse dipole
  field for several impeller rotation rates. The thick line is an
  example Poisson distribution. The thin line represents the energy
  distribution if the magnetic fluctuations were Gaussian.}
\label{fig:Energy_PDFs}
\end{figure}

Table~\ref{tab:burst} reveals that the standard deviation of the
energy in the intermittent transverse dipole field is approximately
equal to its mean value, a characteristic of a Poisson probability
distribution~\cite{Landau_and_Lifshitz}. Assuming each excitation can
be treated as a rare random event, the probability distribution of the
magnetic field energy can be determined heuristically. The probability
of measuring $n$ bursts in time $t$ is given by $P(t) = (f_b t)^n
e^{-f_b t} / n!$ where $f_b$ is the average rate of bursts. The
average growth of the magnetic field over time $t$ during a burst is
$\Delta B = B_0 e^{\lambda t}$, where $B_0$ is the average strength of
the initial seed field. The resulting gain in the magnetic field
energy per unit volume is $\Delta E = \Delta B^2 / 2 \mu_0 = ( B_0^2 /
2\mu_0 ) \exp(2 \lambda t)$ and so $t = \log( \Delta E / E_0 ) / 2
\lambda$ where $E_0 = B_0^2/ 2\mu_0$. Substituting the time in terms
of $\Delta E$ into the Poisson distribution yields a log-Poisson
distribution for the probability density of $\Delta E$:
\begin{equation}
P(\Delta E) = \frac{1}{n!} \left[ \frac{f_b}{2\lambda}
      \ln\left(\frac{\Delta E}{E_0} \right) \right]^n e^
      {-(f_b/2\lambda) \ln\left(\Delta E/E_0 \right)}.
\end{equation}
The probability distributions of the transverse dipole energy are shown
in Fig.~\ref{fig:Energy_PDFs}. The distributions with large numbers of
bursts tend to have significantly more high energy fluctuations than
is expected from Gaussian fluctuations. The overall
invariance of the distributions as the impeller rotation rate is
increased demonstrates that the increased frequency of bursts is
offset by their shortened duration.

The results presented demonstrate how turbulence in a simply-connected
geometry changes the onset conditions of the dynamo.  Rather than the
smooth transition from damped to growing fields predicted by either
kinematic or mean field dynamo theory, the transition is characterized
by intermittent magnetic field bursts which may be relevant to some
dynamo models~\cite{Ko.APJ.1989}. Although sustained growth is
not yet observed, the transient excitation demonstrates the
intermittent characteristics of a turbulent dynamo.

The authors would like to thank S.\,A.~Boldyrev and E.\,G.~Zweibel for
their helpful suggestions and C.\,A.~Parada for his assistance in the
experiments. This work is funded by the Department of Energy, the
National Science Foundation, and the David and Lucille Packard
Foundation.

\end{document}